\documentclass[11pt,twoside]{article}
\usepackage{asp2004}
\usepackage{psfig}
\usepackage{epsf}
\usepackage{graphics}
\usepackage{graphicx}
\usepackage{lscape}
\begin{document}
\title{Time-resolved observations of the short period CV SDSS~J123813.73-033933.0}
\author{S.V. Zharikov$^1$, G. H. Tovmassian$^1$, V. Neustroev$^2$, R. Michel$^1$, R.~Napiwotzki$^3$}
\affil{$^1$ OAN, IA UNAM, Ensenada, BC, M\'exico \\
$^2$ National University of Ireland, Galway, Ireland \\
$^3$ University of Leicester, England }

\begin{abstract}
We  present simultaneous spectral and photometric  observations of
SDSS~J123813.73-033933.0. From H$_\alpha$ radial  velocity
measurements we determined the orbital period  of the  system to
be $0.05592\pm0.00002$  days (80.53 min). The spectrum shows
double Balmer  emission lines flanked by strong, broad absorption,
indicating a  dominant contribution from the white dwarf. The
photometric light curve  shows   complex  variability. The system
undergoes  cyclic brightening up to 0.4  mag which are
semi-periodical on short time scales with periods of the order of
7-12 hours. We also detect 40.25 min variability (\~0.15 mag) in
the  light curve, that corresponds to half the orbital period. Its
amplitude increases with the cyclic brightening of the system.
\end{abstract}

\section{Observations and Data Analysis}
The discovery  of SDSS J123813.73-033933.0  (SDSS1238) was
announced in the second CV release of the SDSS (Szkody et al.
2003). The optical spectrum shows a  blue continuum with broad
absorption around double peaked Balmer emission lines. From its
highly doubled lines, the inclination of the system was suspected
to be high. Here we present simultaneous time resolved spectral
and photometric  observations  of SDSS1238 obtained during April
and May 2004 at SPM  Observatory, Mexico. Spectral observations
were  done in four consecutive nights in April and one night in
May 2004 with a total coverage of ~21h. Differential photometry
was done in V and R bands in  long time series (3 nights in April
and 4 nights in May).
\begin{figure}[!ht]
\plottwo{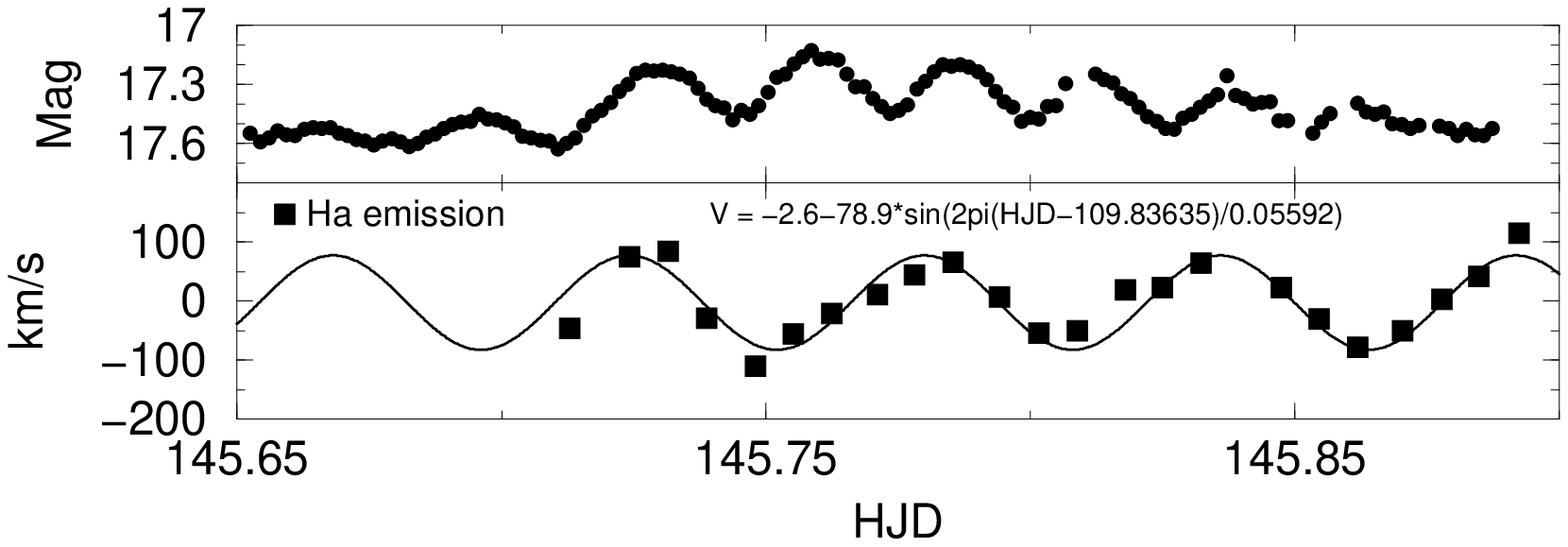}{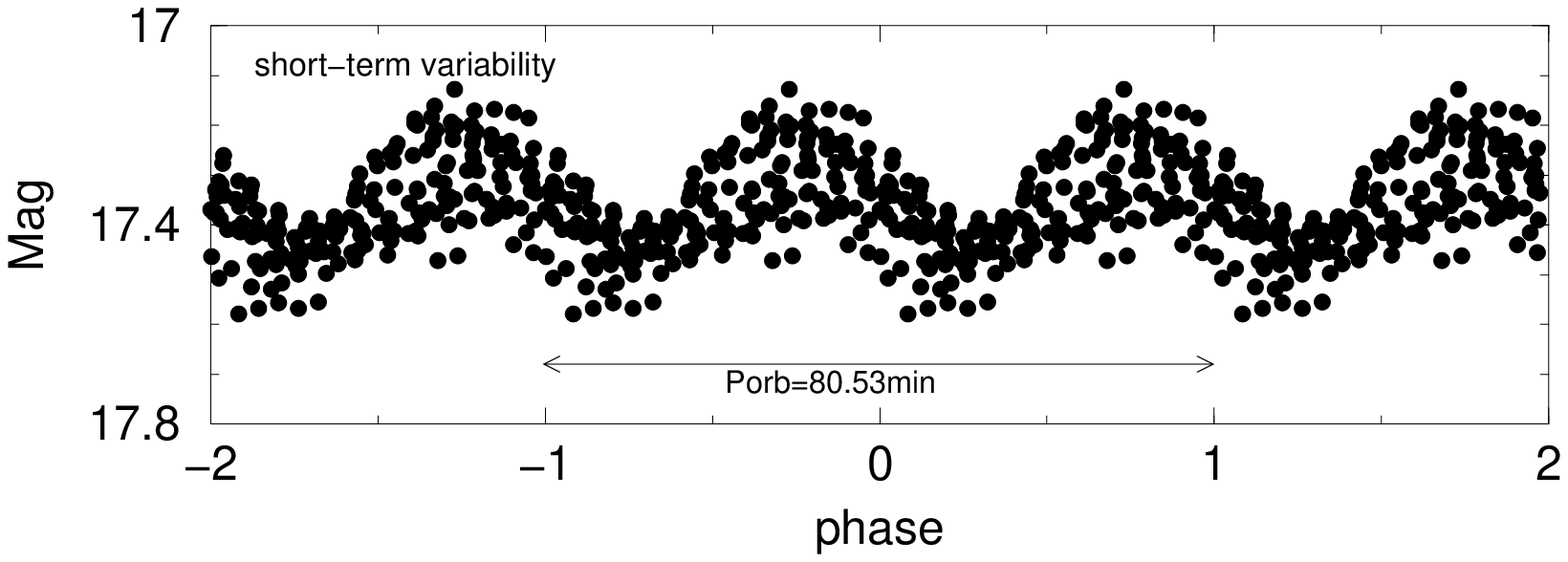} \caption{Left: light
curve and H$_\alpha$ RV curve of SDSS1238 on May 19,
 2004. Right: The short term variability folded with half the orbital period.}
\end{figure}
\begin{figure}
\plottwo{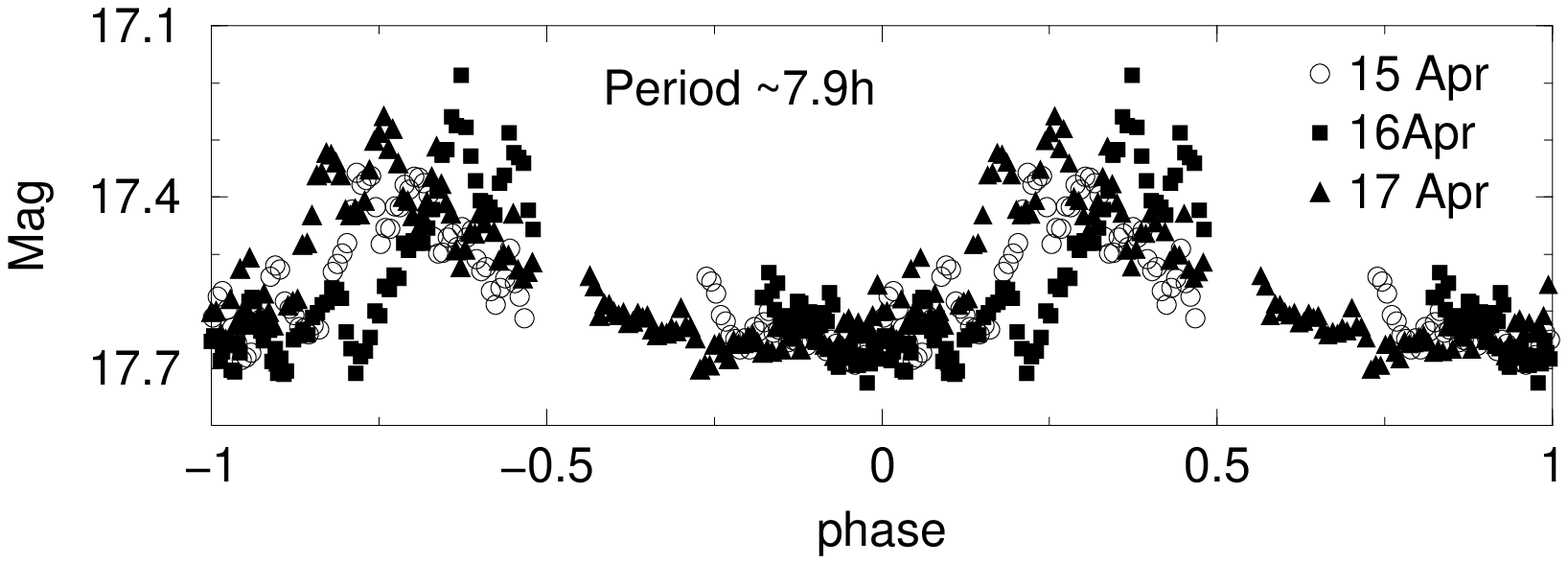}{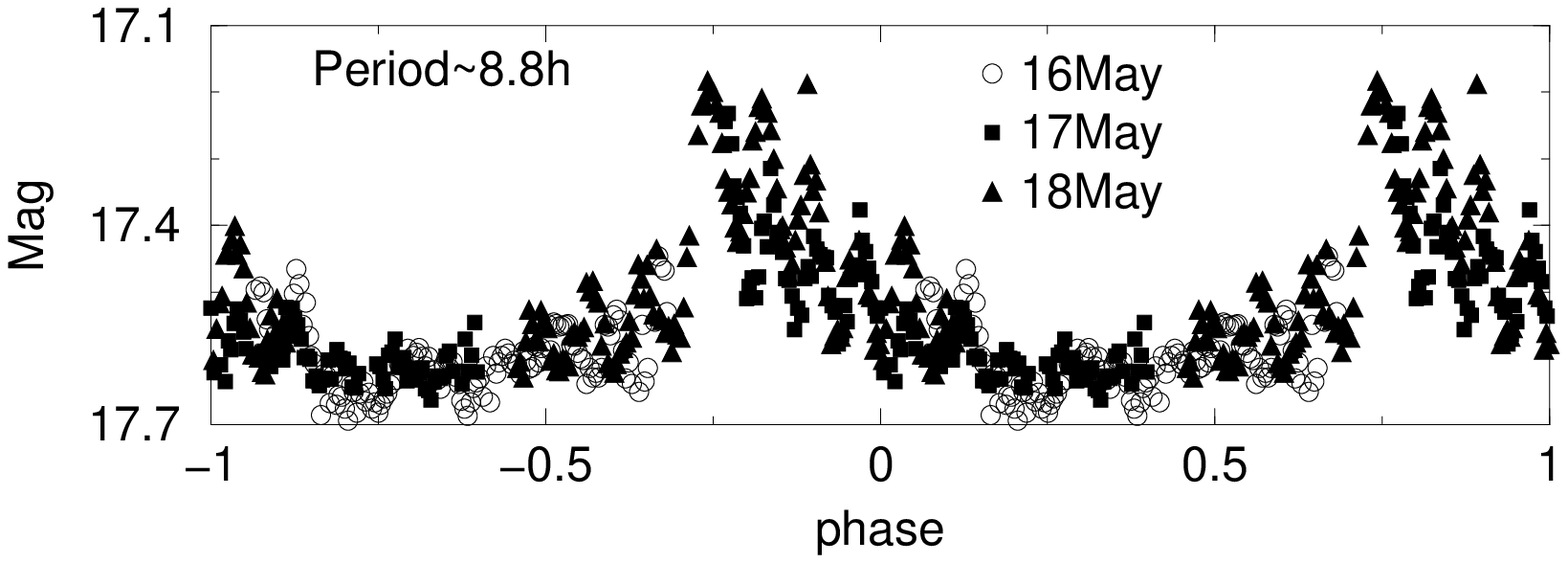} \caption{The
long-term variability of SDSS1238 in April (left) and May (right)
observational runs.}
\end{figure}
The spectrum  of SDSS1238 shows features characteristic of
short-period, low mass transfer, high inclination CVs like WZ Sge
\& RZ Leo. The fit by a WD model of the Balmer absorption features
shows that the Balmer decrement is probably produced by a WD
rather than an accretion disc. Best fit model solutions correspond
to high gravity, indicating a high WD mass ($0.8 -1M_{\sun}$). The
power spectrum of RV variations of the H$_\alpha$ emission line
peaks at a 17.88218 day$^{-1}$, corresponding to an orbital period
$P_{orb}=0.05592\pm0.00002$ days, about 4 min longer than that
recently reported by Szkody et al. (2003). The period analysis
shows that the period and phase of the absorption feature coincide
with the emission component. Photometrically, SDSS1238 shows two
types of variability: a long-term one (7-12h) with an amplitude
$\sim$0.45 mag and a short-term one ($\sim$0.35 mag) at half the
orbital period, more visible in the bright part of the long-term
variability. The maxima of the short-time photometric variability
are at orbital phases 0.375 and 0.875. Long-term variability (LTV)
detected during April showed a distinct periodic (~7.9h) pattern.
A month later, we confirmed the existence of a LTV but found that
its period changed from about 9h to about 12h. The Doppler maps
constructed with the bright and faint phases of the LTV cycle show
different states of the accretion flows in the system. The
photometric behavior of the system is complex and so far
incomprehensible. The half-orbital (spectral) period photometric
variations, usually attributed to the elliptically distorted
secondary, do not make sense in this case when coupled with the
absence of any sign of the secondary in the spectrum, the fact
that they become significantly stronger at brighter phases of LTV
and their phasing with the spectral data. Another rather unusual
explanation includes radiation from {\it symmetric} hot spots in
the disc or at the surface of WD. A double-hump modulation of the
light curve has been observed by Rogoziecki \&
Schwarzenberg-Czerny (2001) in a quiescent state of the short
period (0.05827d) WZ Sge -type system WX Cet in Oct. 1998 and by
Kafka \& Honeycutt, (2003) in the AM Her system QQ Vul at some
observational runs. We interpret the LTV and the difference of the
Doppler maps in bright and faint parts of the light curve of
SDSS1238 as a result of changes of the accretion rate in the
system with the quasi-period 7-12h. This quasi-periodic
brightening is by no means outburst-like and the intervals are
much shorter than the shortest known cycles (3-4 days) of  ER UMa
type outbursts. One can speculate that increased mass-transfer at
bright phases of LTV increases the intensity of the spots causing
the half-orbital period short term variability. The system
certainly requires close monitoring.

\acknowledgments
The work was supported by DGAPA IN110002 project.

\end{document}